\documentclass[superscriptaddress,groupedaddress,nofootnoteinbib,12pt]{article} 
\usepackage{color}
\usepackage{graphicx}  
\usepackage{dcolumn}   
\usepackage{bm}       
\usepackage{amssymb}  
\usepackage{amsmath}
\usepackage{sectsty}
\usepackage{colortbl}
\usepackage{latexsym}
\usepackage{float}
\usepackage{ifthen}
\usepackage{caption,subfig}
\usepackage{enumerate}
\usepackage{url}
\usepackage{caption,subfig}
\usepackage{feynmf}	
\usepackage{jcappub}

\def\be{\begin{equation}}
\def\ee{\end{equation}}
\def\ba{\begin{eqnarray}}
\def\ea{\end{eqnarray}}
\def\beq{\begin{eqnarray}}
\def\eeq{\end{eqnarray}}

\def\mpl{M_{\rm Pl}}

\def\E{\mathcal{E}}

\def\d{\mathrm{d}}
\def\p{{\cal P}}
\def\S{\Sigma}

\def\L*{{\cal L}_*}
\def\L{\mathcal{L}}
\def\({\left(}
\def\){\right)}
\def\ie{{\it i.e. }}
\def\etal{{\it et.al. }}
\def\nn{\nonumber}
\def\p{\partial}
\def\mn{_{\mu \nu}}
\def\ab{_{ab}}
\def\stu{St\"uckelberg }
\def\p{\partial}

\def\<{\langle}
\def\>{\rangle}
\def\Ein{\hat{\mathcal{E}}}

\def\cs2{c_{s}^{2}}

 \def\p{\partial}

\def\be{\begin{equation}}
\def\ee{\end{equation}}
\def\ba{\begin{eqnarray}}
\def\ea{\end{eqnarray}}
\def\beq{\begin{eqnarray}}
\def\eeq{\end{eqnarray}}

\def\mpl{M_{\rm Pl}}

\def\E{\mathcal{E}}

\def\d{\mathrm{d}}
\def\p{{\cal P}}
\def\S{\Sigma}

\def\L*{{\cal L}_*}
\def\L{\mathcal{L}}
\def\({\left(}
\def\){\right)}
\def\ie{{\it i.e. }}
\def\etal{{\it et.al. }}
\def\nn{\nonumber}
\def\p{\partial}
\def\mn{_{\mu \nu}}
\def\stu{St\"uckelberg }
\def\p{\partial}

\def\<{\langle}
\def\>{\rangle}
\def\Ein{\hat{\mathcal{E}}}

\def\intk{\int\frac{\d^4k}{(2\pi)^4}}

\def\propm{\(k^2+m^2\)}
\def\Tr{{\rm Tr}}
 \def\be   {\begin{equation}}   \def\ee   {\end{equation}}

 \def\ba  {\begin{eqnarray}}   \def\ea  {\end{eqnarray}}

\renewcommand{\thesubsection}{\arabic{section}.\arabic{subsection}}

\newcommand{\para}[1]{\par\vspace{2mm}\noindent\emph{{#1}}.---}

\begin{document}

\title{Quantum Corrections in Massive Gravity}

\author{Claudia de Rham,$^{a}$ Lavinia Heisenberg$^{a,b}$ and Raquel H. Ribeiro$^{a}$}
\affiliation{$^{a}$Department of Physics, Case Western Reserve University, \\
10900 Euclid Ave, Cleveland, OH 44106, USA}
\affiliation{$^{b}$D\'epartment de Physique  Th\'eorique and Center for Astroparticle Physics,\\
Universit\'e de Gen\`eve, 24 Quai E. Ansermet, CH-1211  Gen\`eve, Switzerland}

	\emailAdd{Claudia.deRham@case.edu}
	\emailAdd{Lavinia.Heisenberg@unige.ch}
	\emailAdd{RaquelHRibeiro@case.edu}

\abstract{We compute the one-loop quantum corrections to the potential of ghost-free massive gravity. We show how the mass of external matter fields contribute to the running of the cosmological constant, but do not change the ghost-free structure of the massive gravity potential at one-loop.  When considering gravitons running in the loops, we show how the structure of the potential gets destabilized at the quantum level, but in a way which would never involve a ghost with a mass smaller than the Planck scale. This is done by explicitly computing the one-loop effective action and supplementing it with the Vainshtein mechanism. We conclude that to one-loop order the special mass structure of ghost-free massive gravity is technically natural.
}

\maketitle

\section{Introduction}

Over the past century a considerable amount of effort has been devoted to
understanding gravity. With the observational evidence of the recent
accelerated expansion of the Universe \cite{Perlmutter:1997zf,Riess:1998cb,Tonry:2003zg}, this quest has become more urgent
because it might be an indication that we do not understand
gravity at the largest distance scales.
Another striking puzzle in cosmology is the cosmological
constant problem \cite{Weinberg:1988cp}, for which the discrepancy between field theory predictions
and  observations is of many orders of magnitude.

A similar disparity arises in the standard model of particle physics, the hierarchy problem. This is the problem of why the Higgs mass is so small relative to the Planck/unification scales.
These hierarchies are puzzling as they do not seem to be protected without the help of new physics. {\it Technically natural} tunings on the other hand are less of an issue and are common within the standard model. For example, the electron mass, $m_{\rm e}$, is hierarchically smaller than the electroweak scale. In the limit $m_{\rm e}\to0$ there is an enhancement of the
symmetry of the system, due to the recovery of chiral symmetry. The existence of this symmetry in the massless limit is enough to protect the electron mass from receiving large quantum corrections thanks to 
the  \emph{'t Hooft naturalness} argument \cite{tHooft:1979bh,Dimopoulos:1979es}.
Therefore quantum corrections will only give rise to a renormalization of the electron mass proportional to $m_{\rm e}$ itself, and thus the hierarchy between the electron mass and the electroweak scale is technically natural.

In the case of the cosmological constant,  $\Lambda_{\rm CC}$,  there is no symmetry recovered in the limit $\Lambda_{\rm CC}\to 0$.
This is because the (anti-)de Sitter and Poincar\'{e} groups have the same number of generators. Its tuning to a small scale compared to the Higgs mass for instance is therefore unnatural in the 't Hooft sense.
In the context of General Relativity (GR) it has been very hard to address this problem, which has motivated the search for local modifications of gravity in the infra-red.
One possibility is to introduce a technically natural small scale which could account for the late time acceleration of the Universe.
This can be achieved by giving the mediator of gravitational interactions, the graviton, a small mass of order of the Hubble constant today.

However, such a modification usually comes with a number of pathologies \cite{ArkaniHamed:2002sp,Creminelli:2005qk}, most notably the presence of the so-called Boulware--Deser ghost, \cite{Boulware:1973my}. In recent years we have seen considerable progress in formulating a well-defined, non-linear theory of massive gravity, which has no propagating ghost degrees of freedom \cite{Nibbelink:2006sz,Gabadadze:2009ja,deRham:2010gu,deRham:2012az}. Such a theory has been proposed in Refs.~\cite{deRham:2010ik,deRham:2010kj}, which extends the Fierz--Pauli action \cite{Fierz:1939ix,VanNieuwenhuizen:1973fi} and relies on a very specific structure of a 2-parameter family interaction potential. The proof of the absence of ghosts has now been generalized to a multitude of languages and formalisms, see for instance 
Refs.~\cite{Hassan:2011hr,deRham:2011rn,Hassan:2011ea,deRham:2011qq,Mirbabayi:2011aa,Golovnev:2011aa,Hassan:2012qv,Hinterbichler:2012cn,Kluson:2012wf,Deffayet:2012zc}. 
Despite these recent developments, the stability of massive gravity against quantum corrections
remains an open question, 
especially when including interactions with other massive matter fields. This question is not only tied to the smallness of the graviton mass when taking quantum corrections into account; it also raises the issue of whether the quantum corrections are capable of detuning/destabilizing the structure of the potential of massive gravity. Since it is the special nature of this potential which prevents the presence of the Boulware--Deser ghost, a detuning would reintroduce the ghost at some energy scale.

It is therefore crucial to address the question of the quantum stability of massive gravity, and we shall take 
some first steps in this direction. We will also clarify the role played by the Vainshtein mechanism \cite{Vainshtein:1972sx,Deffayet:2001uk, Berezhiani:2013dw, Berezhiani:2013dca} at the level of the quantum corrections, given its importance in enhancing the strongly coupled interactions and curing the classical discontinuity problem \cite{Kogan:2000uy,Porrati:2000cp,Dilkes:2001av,Babichev:2009jt} which arises when taking the massless limit ---this is known as the van Dam--Veltman--Zakharov (vDVZ) discontinuity \cite{vanDam:1970vg,Zakharov:1970cc}. 
Even in cases where the vDVZ discontinuity seems absent at the classical level, it has been known to reappear at the quantum level, either through anomalies in the case of spin 3/2, \cite{Duff:2002sm} or directly in the loops for massive spin-2 fields on AdS, \cite{Dilkes:2001av}.
This paper extends the findings of Ref.~\cite{deRham:2012ew} which discussed the quantum corrections in a class of Galileon theories corresponding to massive gravity in the \emph{decoupling limit}.\footnote{By decoupling limit we mean
                                                                 taking $m\rightarrow 0 $  and
                                                                 $\mpl\rightarrow \infty$ simultaneously,
                                                                 whilst keeping
                                                                 the energy scale $\Lambda_3\equiv
                                                                 (m^2\mpl)^{1/3}$ fixed.}

In this paper we study how and at what scale the specific structure of the graviton gets detuned,
and compare our conclusions to previous results.\footnote{Ref.~\cite{Capper:1973bk} is an example of an early work which discussed the quantum corrections in GR in a similar way this paper explores the radiative corrections to massive gravity.}
We emphasize we are \emph{not} addressing the old cosmological constant problem,
since $\sqrt{-g}\,\Lambda_{\rm CC}$ is one of the allowed ghost-free potential terms for the graviton,
and it is therefore harmless for our discussion.

For the quantum stability analysis, we focus on quantum corrections arising at one-loop only, and leave the extension to higher order loops to a further study, where the mixing between graviton and scalar propagators within loops can occur \cite{2loops}. In particular this implies that either gravitons or matter field are running in the loops, but not \emph{both} simultaneously.
Furthermore, by working in dimensional regularization, we discard any measure issues in the path integral related to field redefinitions which show up in power law divergences. We thus focus on logarithmic running results which are independent of this measure factor---in the language of field theory we 
concentrate on the runnings of the couplings.

 Moreover, our aim 
 is to study the stability of the graviton {\it potential} against quantum corrections, rather than the whole gravity action. As a result it is sufficient to address the diagrams for which the external graviton legs have zero momenta (\ie, we focus on the IR limit of the runnings). 
This
approach is complementary to the work developed 
by Buchbinder \etal in \cite{Buchbinder:2012wb} who used the Schwinger--DeWitt expansion of the one-loop 
effective action. This method allows one to obtain the 
Seeley--DeWitt coefficients associated with
the curvature invariants generated
by quantum corrections (see also Ref. \cite{Buchbinder:2007xq}).
Our approach differs in two ways. First, we 
introduce a covariant coupling to the matter sector 
and obtain the quantum corrections generated by matter loops. 
And second, we go beyond the minimal model
investigated by Ref. \cite{Buchbinder:2012wb} and study the quantum corrections
to the full potential. As a by product, we do not focus on the 
radiatively generated curvature terms since these would also arise in GR
and would therefore not be exclusive of theories of massive gravity.

For convenience, when computing the one-loop effective action, we consider a background configuration for the metric which is spacetime independent.  
 Furthermore,  we work in Euclidean space, and use the mostly $+$ signature convention. Massive gravity on a Minkowski reference metric is thus mapped to massive gravity on a flat Euclidean reference metric $\delta_{ab}$ in Euclidean space.
Finally, we use units for which $\hbar=1$. 

The remainder of this paper is organized as follows. In section \ref{sec:dRGT} we review the tree level ghost-free covariant non-linear theory of massive gravity and briefly discuss the quantum corrections in the decoupling limit. We then move on to the full theory and concentrate in section \ref{sec:MatterLoops} on the one-loop contributions arising from the coupling to external matter fields. We show they only imply a running of the cosmological constant and of no other potential terms for the graviton.

In section \ref{sec:GravitonLoops} we
discuss the one-loop contributions from the gravitons themselves, and show that whilst these destabilize the special structure of the potential, 
this detuning is irrelevant below the Planck scale. 
We then push the analysis further in section \ref{sec:1PI} and show that even if the background configuration is large, 
 as should be the case for the Vainshtein mechanism to work \cite{Vainshtein:1972sx}, this will redress the one-loop effective action in such a way that the detuning remains irrelevant below the Planck scale. We end by summarizing our results and presenting some open questions in section~\ref{sec:conclusion}. The appendices collect extra material:
appendix \ref{sec:appendixDimReg} states
our conventions for the dimensional regularization scheme, whilst appendix \ref{app:matter}
presents all the details of the calculation involving matter loops only.

\section{Ghost-free massive gravity}
\label{sec:dRGT}
In this section we review the ghost-free interactions in the theory of massive gravity.
We also summarize the findings of Ref.~\cite{deRham:2012ew}
where the quantum stability of the classical theory in the decoupling limit was studied.
This will be our starting point to investigate the way quantum corrections affect the general structure of the potential.

The presence of a square-root in the ghost-free realization of massive gravity makes its expression much more natural in the vielbein language \cite{Nibbelink:2006sz,Chamseddine:2011mu,Hinterbichler:2012cn,Deffayet:2012zc} (see also
Refs.~\cite{Gabadadze:2013ria,Ondo:2013wka}). In the vielbein formalism, the ghost-free potential is polynomial and at most quartic in the vielbein fields.
To make use of this natural formulation, we will work throughout this paper in a `symmetric-vielbein inspired language' where the metric is given by
\ba
\label{vielbeinForm}
g_{ab}=\left(\bar \gamma_{ab}+\frac{h_{ab}}{\mpl}\right)^2
\equiv
\left(\bar \gamma_{ac}+\frac{h_{ac}}{\mpl}\right)\left(\bar \gamma_{db}+\frac{h_{db}}{\mpl}\right)\delta^{cd}\,,
\ea
where $\bar g_{ab}=\bar\gamma\ab^2=\bar\gamma_{ac}\bar\gamma_{bd}\delta^{cd}$ is the background metric, 
and $h\ab$ plays the role of the fluctuations. We stress that the background metric $\bar g_{ab}$ need not be flat, even though the reference metric $f\ab$ will be taken to be flat throughout this study, $f\ab=\delta\ab$.\footnote{In the Euclidean version of massive gravity both the dynamical metric $g\mn$ and the reference metric $f\mn$ have to be `Euclideanized', $g\mn\to g\ab$ and $f\mn=\eta\mn \to \delta\ab$.}

In this language, when working around a flat background metric, $\bar \gamma_{ab}=\delta_{ab}$, the normal fluctuations about flat space are expressed in terms of $h$ as
\ba
g_{ab}-\delta_{ab}=\frac{2}{\mpl}h_{ab}+\frac{1}{\mpl^2}h_{ac}h_{bd}\delta^{cd}\,.
\label{eq:defmetric}
\ea
The conversion from $g_{ab}$ to $h_{ab}$ is a field redefinition that will contribute a measure term in the path integral. This generates power law divergent corrections to the action which, since we will work in dimensional regularization, can be ignored. This reflects the fact that the physics is independent of such field redefinitions, and only the logarithmic runnings are
physically meaningful for the purposes of our study.

\subsection{Ghost-free potential}\label{potentialdRGT}

A two-parameter family of potentials for the graviton has been proposed in
Refs.~\cite{deRham:2010ik,deRham:2010kj}. These potential terms were built in such a way so as to remove
higher derivative terms in the \stu fields which would otherwise induce a propagating ghost degree of freedom \cite{Boulware:1973my,Deffayet:2005ys}.
Consider a graviton of mass $m$
\ba
\label{eq:full}
\mathcal L_{\rm mGR}=-\frac{\mpl^2}{2}\sqrt{g}\left(R-\frac{m^2}{4}\mathcal U(g,H)\right) \,,
\ea
where the overall minus sign arises after Wick rotation to Euclidean space.\footnote{As already mentioned, for the purposes of computing loop corrections,
it is  more convenient to work with the flat Euclidean reference
metric, $\delta_{ab}$, after performing a Wick rotation $t \to -i \tau$, where $\tau$ is the Euclidean time.}
In this action $\mathcal U$ is the potential and the tensor $H\ab$ is constructed out of the metric $g\ab$ and the four \stu fields $\Phi^a$ \cite{Siegel:1993sk,ArkaniHamed:2002sp},
\ba
 H\ab=g\ab-\delta_{cd}\p_a \Phi^c \p_b \Phi^d.
 \label{eq:defHmunu}
 \ea
For the purpose of this section, we can work with a flat background metric $\bar g\ab=\delta\ab$.

One can split the $\Phi^a$'s into a helicity-0 and-1 modes. For the sake of this argument, it is sufficient to focus on the helicity-0 mode and set the helicity-1 to zero, so that $\Phi^a=(x^a-\delta^{ab}\partial_b\hat{\pi})$, where $\pi=\Lambda_3^3 \,\hat \pi$ is the helicity-0 mode, and
we recall that $\Lambda_3^3=m^2 \mpl$. Then, in terms of the helicity-0 and -2 modes, the tensor $H\ab$ becomes
\ba
H\ab=2 \hat{h}\ab+ \hat{h}_a^{\ c} \hat{h}_{bc}+2\hat{\Pi}\ab-\delta^{cd}\hat{\Pi}_{ac}\hat{\Pi}_{bd}\ .
\label{eq:defH}
\ea
In this notation $\hat{\Pi}\ab \equiv \partial_a \partial_b \pi/\Lambda_3^3$, and $h\ab \equiv \mpl \hat h\ab $ is the canonically normalized helicity-2 mode. Indices are lowered and raised with respect to the flat background metric, $\delta\ab$.

The \stu fields have been introduced so as to restore diffeomorphism invariance. We are now free to set a gauge for the metric or the \stu fields, and can choose in particular the unitary gauge, where $\Phi^a(x)=x^a$. In that gauge,  the tensor $H\mn$ is simply given by $H\ab=2 \hat{h}\ab +\hat{h}_a^{\ c} \hat{h}_{cb}$. In terms of these `vielbein-inspired' perturbations, the ghost-free potential becomes polynomial in $h\ab$,
\ba
\mathcal{L}=-\dfrac{\mpl^2}{2} \sqrt{g}R
-\dfrac{1}{4} \mpl^2 m^2
\sum_{n=2}^4{\frac{1}{\mpl^n }\tilde{\alpha}_n\, \tilde{\mathcal{U}}_n[H]}\, .
\label{eq:Lvielbein}
\ea
In unitary gauge the potential above is fully defined by
\ba
\label{eq:U2h}
\tilde{\mathcal{U}}_2[h] &=&  \E^{abcd}\E^{a'b'}_{\ \ \ cd} h_{aa'}h_{bb'}= -2 \(h\ab h_{cd}-h_{ac}h_{bd}\)\delta^{ac}\delta^{bd}\
\\
\label{eq:U3h}
\tilde{\mathcal{U}}_3[h] &=&  \E^{abcd}\E^{a'b'c'}_{\ \ \ \ \ d} \
 h_{aa'}h_{bb'}h_{cc'}\\
\label{eq:U4h}
\tilde{\mathcal{U}}_4[h] &=&  \E^{abcd}\E^{a'b'c'd'} \
 h_{aa'}h_{bb'}h_{cc'}h_{dd'}\,,
\ea
where $\E^{abcd}$ represents the fully antisymmetric Levi-Cevita {\it symbol} (and not tensor, so in this language $\E^{abc}_{\ \ \ d}=\delta_{dd'}\E^{abcd'}$, for example,
 carries no information about the metric).

The first coefficient 
is fixed, $\tilde{\alpha}_2=1$, whereas the two others are free.
They relate to the two free coefficients of Ref.~\cite{deRham:2010kj},  as
$\tilde{\alpha}_3=-2(1+\alpha_3)$ and
$\tilde{\alpha}_4=-2(\alpha_3+\alpha_4)-1$
(where $\alpha_3$ and $\alpha_4$ are respectively the coefficients of
the potential $\mathcal{U}_3$ and $\mathcal{U}_4$ in that language).\footnote{As in Ref.~\cite{deRham:2010kj}, this 2-parameter family of potential is the one for which there is no cosmological constant nor tadpole.}
The absence of ghost-like pathologies is tied to the fact that, when expressed in terms of $\pi$ uniquely, \eqref{eq:U2h}--\eqref{eq:U4h} are total derivatives.

\subsection{Quantum corrections in the decoupling limit}

Before moving on to computing the quantum corrections to the potential arising from matter loops, we first review the radiative corrections within the decoupling limit as derived in Ref. \cite{deRham:2012ew}. In this case the helicities-2 and -0 of the graviton decouple from each other and become accessible separately, thereby creating a framework benefiting the visibility of the most important physical properties of the theory. In this limit, the usual helicity-2 mode of gravity can be treated linearly while the helicity-0 mode still contains non-linear interactions. The decoupling limit can be written in the compact form \cite{deRham:2010ik,deRham:2010kj}
\begin{equation}
\mathcal{L}=h^{\mu\nu}\Ein^{\alpha\beta}_{\ \ \mu\nu} h_{\alpha\beta}-
 h^{\mu\nu}\sum_{n=1}^3 \frac{a_{n}}{\Lambda^{3(n-1)}_3} X^{(n)}_{\mu\nu}\!\left(\Pi\right),
\label{lagr1}
\end{equation}
where the Lichnerowicz operator, $\Ein^{\alpha\beta}_{\ \ \mu\nu}$,
acts on the metric perturbations as follows
\ba
\Ein^{\alpha\beta}_{\ \ \mu \nu} h_{\alpha\beta}=-\frac 12 \(\Box h\mn-2\p_\alpha \p_{(\mu}h^\alpha_{\nu)}+\p_\mu\p_\nu h-\delta\mn (\Box h-\p_\alpha\p_\beta h^{\alpha\beta})\)\,.
\label{eq:Lichnerowicz}
\ea
 Here, $h_{(ab)}\equiv\frac{1}{2}(h_{ab}+h_{ba})$, and one can always set $a_1=1$ while the two other dimensionless coefficients $a_{2,3}$ are related to the two free parameters $\tilde \alpha_{2,3}$.
 The three tensors $X_{\mu\nu}^{(n)}$ represent the interactions of order $n$ in $\pi$ through $\Pi_{\mu\nu}$ between the helicities-2 and -0 modes
\begin{eqnarray}
X^{(1)}_{\mu\nu}\left(\Pi\right)&=&{\mathcal{E}_{\mu}}^{\alpha\rho\sigma}
{{\mathcal{E}_\nu}^{\beta}}_{\rho\sigma}\Pi_{\alpha\beta}\ , \quad  \nonumber \\
X^{(2)}_{\mu\nu}\left(\Pi\right)&=&{\mathcal{E}_{\mu}}^{\alpha\rho\gamma}
{{\mathcal{E}_\nu}^{\beta\sigma}}_{\gamma}\Pi_{\alpha\beta}
\Pi_{\rho\sigma} \ ,\ \ \textrm{and} \nonumber \\
X^{(3)}_{\mu\nu}\left(\Pi\right)&=&{\mathcal{E}_{\mu}}^{\alpha\rho\gamma}
{{\mathcal{E}_\nu}^{\beta\sigma\delta}}\Pi_{\alpha\beta}
\Pi_{\rho\sigma}\Pi_{\gamma\delta}\ .
\label{Xs}
\end{eqnarray}
The natural question that arises is whether the parameter $a_n$ and 
the energy scale $\Lambda_3$  are radiatively stable.
Using the antisymmetric structure of the interactions \eqref{Xs},
one can roughly follow the same non-renormalization argument of Galileon theories \cite{Nicolis:2004qq,Luty:2003vm,Nicolis:2008in}
 to show that $a_{2}$ and $a_3$ do not get renormalized
within the decoupling limit of the theory.\footnote{The absence of the ghost in these theories is tightly related to the antisymmetric nature of their interactions, which in turn guarantees their non-renormalization. The same reasoning applies to the construction of the Lovelock invariants. For example, in linearized GR, linearized diffemorphism tells us that the kinetic term can be written using the antisymmetric Levi--Cevita symbols as $\mathcal{E}^{\mu\nu\alpha\beta}\mathcal{E}^{\mu'\nu'}_{\phantom{\mu'\nu'}\alpha\beta}R_{\mu\nu\mu'\nu'}$, which ensures the non-renormalization.  Notice that gauge invariance alone would still allow for a renormalization of the overall factor of the linearized Einstein--Hilbert term, which does not occur in the decoupling limit.}
The key point is that any external particle
attached to a diagram has at least two derivatives acting on it. This in turn implies that the operators generated are all of the form $(\p^2 \pi)^{n_1}(\p^2 h)^{n_2}$, $n_1, n_2 \in \mathbb{N}$, and so are not of the same class as the original operators. This means that $a_2$ and $a_3$ are not renormalized. Furthermore, the new operators that appear in the 1PI are suppressed by higher powers of derivatives.

To be more precise, any external particle contracted with a field with two derivatives in a vertex contributes to a two-derivatives operator acting on this external
particle---this is the trivial case. 
 On the other hand, if we contract the external particles with fields without derivatives we could in principle generate operators with fewer derivatives. But now the antisymmetric structure of the interactions plays a crucial role.
Take for instance the interaction $V \supseteq h^{\mu\nu} \mathcal{E}_{\mu}^{\;\;\alpha\rho\gamma} \mathcal{E}_{\nu\;\;\;\;\gamma}^{\;\;\beta\sigma}\Pi_{\alpha\beta}\Pi_{\rho\sigma}$, and contract an external helicity-2 particle with momentum  $p_\mu$ with the helicity-2 field coming without derivatives in this vertex; the other two $\pi$-particles run in the loop with momenta $k_\mu$ and $(p+k)_\mu$. The contribution of this vertex gives
\begin{eqnarray}
\mathcal A
 \propto  \int \frac{\mathrm{d}^4 k}{(2\pi)^4} G_k \,
 G_{k+p}\  f^{\mu\nu}\, \mathcal{E}_{\mu}^{\;\;\alpha\rho\gamma} \,
\mathcal{E}_{\nu\;\;\;\;\gamma}^{\;\;\beta\sigma}
\, \, k_\alpha \, k_\beta \, (p+k)_\rho \, (p+k)_\sigma \cdots \,,
\end{eqnarray}
where $G_k= k^{-2}$ is the Feynman massless propagator, and $f^{\mu\nu}$ is the spin-2 polarization tensor. The ellipses denote the remaining terms of the diagram, which are irrelevant for our argument.
The only non-vanishing contribution to the scattering amplitude
 will come in with at least two powers of the external helicity-2 momentum $p_\rho p_\sigma$
\begin{eqnarray}
\mathcal A
 \propto  f^{\mu\nu} \mathcal{E}_{\mu}^{\;\;\alpha\rho\gamma}
\mathcal{E}_{\nu\;\;\;\;\gamma}^{\;\;\beta\sigma}\, \, p_\rho p_\sigma \int\frac{\mathrm{d}^4k}{(2\pi)^4} G_k \, G_{k+p} \,k_\alpha \, k_\beta \cdots \,,
\end{eqnarray}
which in coordinate space corresponds to two derivatives,
as argued above.
The same argument applies to any other vertex, such as $h^{\mu\nu}X_{\mu\nu}^{(3)}(\Pi)$.

This is the essence of the `non-renormalization theorem' in the decoupling limit of massive gravity: there are no quantum corrections to the two parameters $a_{2}$ and $a_3$, nor to the scale $\Lambda_3$. Moreover, the kinetic term of the helicity-2 mode is radiatively stable. We refer the reader to Ref.~\cite{deRham:2012ew} for more details.

\subsection{Propagator in unitary gauge}

In this paper our goal is to go beyond the non-renormalization argument in the decoupling limit reviewed above and investigate the quantum corrections in the full non-linear theory. 
We choose to work 
in the unitary gauge in which the \stu fields vanish and $\Phi^a=x^\alpha\delta^a_{\ \alpha}$. $h\mn$ encodes all the five physical degrees of freedom if it is massive (the two helicity-$\pm2$, the two helicity-$\pm1$ and the helicity-$0$ modes), and only the two helicity-$\pm2$ modes if it is massless.
The Feynman propagator for the massless graviton is given by
\ba
\label{GR_propagator}
G^{\rm (massless)}_{abcd}&=&\langle h_{ab}(x_1) h_{cd}(x_2)\rangle=f^{(0)}_{abcd}  \,
\int \frac{\d^4 k}{(2\pi)^4}
\frac{e^{i k \cdot  \(x_1- x_2 \)}}{k^2},
\ea
in which $x_{1,2}$ are the Euclidean space coordinates, and where the polarization structure is given by
\ba
\label{f}
f^{(0)}_{abcd}=\delta_{a(c}\delta_{bd)}-\frac 12  \delta\ab \delta_{cd}\,.
\ea
Here $\delta_{a(c}\delta_{bd)}\equiv\frac{1}{2} \delta_{ac} \delta_{bd}+\frac{1}{2}\delta_{ad}\delta_{bc}.$
For the massive graviton,  on the other hand, the corresponding Feynman propagator is given by
\ba
\label{MG_propagator}
G^{\rm (massive)}_{abcd}&=&\langle h\ab(x_1) h_{cd}(x_2)\rangle=f^{(m)}_{abcd} \int \frac{\d^4 k}{(2\pi)^4}\frac{e^{i k \cdot \(x_1 - x_2\)}}{k^2+m^2}\,,
\ea
with the polarization structure
\ba
f^{(m)}_{abcd}=\(\tilde \delta_{a(c}\tilde \delta_{bd)}-\frac 13 \tilde \delta\ab \tilde  \delta_{cd}\)
\hspace{15pt}{\rm where}\hspace{15pt}
\tilde \delta\ab=\delta\ab+\frac{k_a k_b}{m^2}\,.
\label{eq:fmassive}
\ea
Notice that the polarization of the massive graviton is no longer proportional to $f^{(0)}_{abcd}$.
Consequently when we take the massless limit, $m\to0$, we do not recover the GR limit, which is at the origin of the vDVZ-discontinuity \cite{vanDam:1970vg,Zakharov:1970cc},
\ba
\lim_{m\to 0}f^{(m)}_{abcd} \ne f^{(0)}_{abcd}\ .
\nn
\ea
At first sight this might be worrisome since on solar system and galactic scales gravity is in very good agreement with GR.
Nevertheless, on these small scales, the effects of massive gravity
can be cloaked by the Vainshtein mechanism \cite{Vainshtein:1972sx}, where the crucial idea is to decouple the additional modes
from the gravitational dynamics via nonlinear interactions of the helicity-0 graviton.

The success of the Vainshtein mechanism relies on derivative interactions, which cause the helicity-0 mode to decouple from matter
on short distances, whilst having observational signatures on larger scales. In this paper we will study how the Vainshtein mechanism acts explicitly at the quantum level, and how the quantum corrections do not diverge in the limit when $m\rightarrow 0$, even though the propagator \eqref{MG_propagator} does.

In the next section we focus on the IR behaviour of the loop corrections. Starting with loops of matter, we will see that the peculiar structure in \eqref{eq:fmassive} has no effect on the computation of the
quantum corrected effective potential at one-loop. This is because at one-loop the 
matter field and the graviton cannot both be simultaneously propagating in the loops if we consider only the contributions to the graviton potential. 
As a result, the quantum corrections are equivalent to those in GR,
as is shown explicitly in section \ref{sec:MatterLoops}.
Only once we start considering loops containing virtual gravitons
will the different polarization and the appearance of the mass in the propagator have an
impact on the  results, as we shall see in section \ref{sec:GravitonLoops}.
Furthermore, the graviton potential induces new vertices which also ought to be considered.

\section{Matter loops}
\label{sec:MatterLoops}

In the previous section, we have reviewed how the `non-renormalization theorem' prevents large quantum corrections from arising in the decoupling limit of massive gravity. Since the coupling to external matter fields is suppressed by the Planck scale these decouple completely
when we take $\mpl\to \infty$.

In this section, we keep the Planck scale, $\mpl$, finite.
We look at the contributions from matter loops and investigate their effect on the structure of the graviton potential. For definiteness, we consider gravity coupled to a scalar field $\chi$ of mass $M$ and study one-loop effects. When focusing on the one-loop 1PI for the graviton potential, there can be no mixing between the graviton and the scalar field inside the loop (this mixing only arises at higher loops, which we discuss in further work \cite{2loops}).
Furthermore,
since we are interested in the corrections to the graviton potential,
we only assume graviton zero momentum for the external legs.
We use dimensional regularization so as to focus on the running of the couplings, which are encoded by the
logarithmic terms.

\subsection{Framework}

Our starting point is the Lagrangian for massive gravity \eqref{eq:full} 
to which we add a real scalar field $\chi$ of mass $M$,
\ba
S=\int \d^4x \(\L_{\rm mGR}+\L_{\rm matter}\)\,,
\ea
with
\ba
\label{Lmatter}
\L_{\rm matter}=\sqrt{g}\(\frac 12 g^{ab}\p_a\chi \p_b \chi+\frac 12 M^2 \chi^2\)\,.
\ea
Note the sign difference owing to the fact that this is the Euclidean action.
The Feynman propagator for the scalar field reads
\ba
\label{scalar_propagator}
G_{\chi}&=&\langle \chi(x_1) \chi(x_2)\rangle=\int \frac{\d^4 k}{(2\pi)^4}\frac{e^{i k . \(x_1 - x_2\)}}{k^2+M^2}\,.
\ea
The mixing between the scalar field and the graviton is encoded in  \eqref{Lmatter} and is highly non-linear.
Before proceeding any further it is convenient to perform the following change of variables for the scalar field
\ba\label{Redefinition}
\chi \to (g)^{-1/4} \psi\,,
\ea
where $g\equiv \det\{g\ab\}$,
so that the matter Lagrangian is now expressed as
\ba
\L_{\rm matter}=\frac 12g^{cd}\(\p_c \psi -\frac 14 \psi g^{ab}\p_c g\ab\)\(\p_d \psi-\frac 14 \psi g^{pq}\p_d g_{pq}\)+\frac 12 M^2 \psi^2 \,.
\ea
Since we will only be considering zero momenta for the external graviton legs, we may neglect the terms of the form $\p g$.\footnote{Such terms were kept in Ref. \cite{Park:2010rp} where the corrections involved higher order curvature
invariants built out of the metric perturbations, but in this study we are only interested in the corrections to the graviton potential and not the higher curvature terms.} As a result, the relevant action for computing the matter loops in given by
\ba
S_{\rm matter}=\int \d^4x \(\frac 12g^{cd}\p_c \psi \p_d \psi +\frac 12 M^2 \psi^2 \)\,.
\ea
In what follows we will compute the one-loop effective action (restricting ourselves to a scalar field in the loops only) and show explicitly that the interactions between the graviton and the scalar field lead to the running of the cosmological constant, but not of the graviton potential.
This comes as no surprise since inside the loops the virtual scalar field has no knowledge of the graviton mass and thus behaves 
in precisely the same way as in GR, leading to a covariant one-loop effective action.
When it comes to the potential, the only operator it can give
rise to which is covariant is the cosmological constant. We show this result explicitly in the one-loop effective action, and then  present it in a perturbative way, which will be more appropriate when dealing with the graviton  loops.

\subsection{One-loop effective action}
\label{subsec:oneloopea}

The one-loop effective action $S_{1,\textrm{eff}}(g\ab, \psi)$ is given by
\ba
e^{- S_{1,\textrm{eff}}(\bar g\ab, \bar \psi)}=\int \mathcal{D}\Psi e^{- \Psi^i  \(S_{ij}(\bar g\ab, \bar \psi)\) \Psi^j}\,,
\ea
where $\Psi_i$ is a placeholder for all the fields, $\Psi_i=\{g\ab, \psi\}$, and $S_{ij}$ is the second derivative of the action with respect to those fields,
\ba
S_{ij}(\bar g\ab, \bar \psi)
\equiv
 \frac{\delta^2 S}{\delta \Psi^i\delta \Psi^j}\Big|_{g\ab=\bar g\ab, \psi=\bar \psi}\,.
\ea
Here $\bar g\ab$ and $\bar{\psi}$ correspond to the background quantities around
which the action for fluctuations is expanded.
Since we are interested in the graviton potential part of the one-loop effective action, we may simply integrate over the scalar field and obtain
\ba
e^{- S^{({\rm matter-loops})}_{1,\textrm{eff}}(\bar g\ab)}=\int \mathcal{D}\psi e^{-\psi \(\frac{\delta^2 S}{\delta^2 \psi}|_{g\ab=\bar g\ab}\)\psi}\,.
\ea
We therefore recover the well-known Coleman--Weinberg effective action,
\ba
S^{({\rm matter-loops})}_{1,\textrm{eff}}(\bar g\ab)= \frac 12 \log \det  \(\frac{\delta^2 S}{\delta^2 \psi}\Big|_{g\ab=\bar g\ab}\)
=\frac 12 \Tr \log \(\frac{\delta^2 S}{\delta^2 \psi}\Big|_{g\ab=\bar g\ab}\)\,.
\ea
Going into Fourier space this leads to
\ba
\L^{({\rm matter-loops})}_{1,\textrm{eff}}(\bar g\ab)&=& \frac 12  \int \frac{\d^4 k}{(2\pi)^4} \log\(\frac 12 \bar g^{ab}k_a k_b+\frac 12 M^2\) \nonumber\\
&=& \frac12 \sqrt{\bar g}\int \frac{\d^4 \tilde k}{(2\pi)^4}  \log\(\frac 12 \delta^{ab}\tilde k_a \tilde k_b+\frac 12 M^2\)\nonumber\\
&\supset& \frac {M^4}{64 \pi^2} \sqrt{\bar g} \log(\mu^2 )\,,
\ea
where $\mu$ is the regularization scale and we restrict our result to the running piece.
From the first to the second equality, we have performed the change of momentum $k_a \rightarrow \tilde k_a$ such that
$g^{ab}k_a k_b = \delta^{ab}\tilde k_a \tilde k_b$.
From this analysis, we see directly that the effect of external matter at one-loop is harmless on the graviton potential. This is no different from GR, since the scalar field running in the loops is unaware of the graviton mass, and the result is covariant by construction. This conclusion is easily understandable in the one-loop effective action
(however, when it comes to graviton loops it will be harder to compute the one-loop effective action non-perturbatively and we will perform a perturbative analysis
instead). For consistency, we
apply a perturbative treatment to the matter fields as well in
section \ref{subsec:renormLambda}.

\para{\bf Higher Loops}
Before moving on to the perturbative argument, we briefly comment on the extension of this result to higher loops. Focusing on matter loops only then additional self-interactions in the matter sector ought to be included. Let us consider, for instance, a $\lambda \chi^3$ coupling. The matter Lagrangian will then include a new operator of the form $\mathcal{L}\supseteq \lambda\sqrt{g}\chi^3=\lambda g^{-1/4}\psi^3$, where $g=\det\{g\ab\}$. 
At $n$-loops, we have $n$ integrals over momentum, and $2(n-1)$ vertices $\lambda g^{-1/4}\psi^3$, so the $n$-loop effective action reads symbolically
\ba
S_n^{({\rm matter-loops})}(\bar g\ab)=\frac{\lambda^{2(n-1)}}{g^{(n-1)/2}}\int \frac{\d^4 k_1 \cdots \d^4 k_n }{(2\pi)^{4n}}\ \mathcal{F}_n\(k_1^2,\cdots,k_n^2, M^2\)\,,
\ea
where $\mathcal{F}_n$ is a scalar function of the different momenta $k_j^2=\bar g^{ab}k_{j a}k_{j b}$.\footnote{Even if different momenta $k_j$ contract one can always reexpress them as functions of $k_j^2$, following a similar procedure to what is presented in appendix~\ref{sec:appendixDimReg}.} As a result one can perform the same change of variables as used previously, $k_j \to \tilde k_j$, with $k_j^2=\delta^{ab}\tilde k_{j a}\tilde k_{j b}\equiv \tilde k_j^2$. This brings $n$ powers of the measure $\sqrt{g}$ down so that the $n$-loop effective action is again precisely proportional to $\sqrt{g}$
\ba
S_n^{({\rm matter-loops})}(\bar g\ab)&=&\frac{\lambda^{2(n-1)}}{g^{(n-1)/2}}g^{n/2}\left[\int \frac{\d^4 \tilde k_1 \cdots \d^4 \tilde k_n}{(2\pi)^{4n}} \ \mathcal{F}_n\(\tilde k_1^2,\cdots,\tilde k_n^2, M^2\)\right] \nonumber\\
&\propto& \sqrt{g} \ \lambda^{2(n-1)}M^{6-2n} \log \mu \,.
\ea
The integral in square brackets is now completely independent of the metric $\bar g\ab$ and the $n$-loop effective action behaves as a cosmological constant. The same result holds for any other matter self-interactions.
Once again this result is not surprising as this corresponds to the only covariant potential term it can be.

\subsection{Perturbative approach}

\label{subsec:renormLambda}

In the previous subsection we have shown how at one-loop external matter fields only affect the cosmological constant 
and no 
other terms in the graviton potential. For consistency we show how this can be seen perturbatively.
As mentioned previously, we use the `vielbein-inspired' metric perturbation about flat space,
as defined in Eq. \eqref{eq:defmetric}.
Including all the interactions between the graviton and the matter field, but ignoring the graviton self-interactions for now, the relevant action is then
\ba
S &= \displaystyle{\int} \d^4 x\   \bigg\{
h^{ab}\left[\Ein\ab^{cd}+\frac 12 m^2 \(\delta_a^c \delta_b^d-\delta\ab\delta^{cd}\)\right]
h_{cd} \nn \\ \vspace*{0.5cm}
& +
\frac{1}{2}\displaystyle{\sum_{n \geq 0}} (-1)^n \,  (n+1) (\hat{h}^{ab})^n
\p_a \psi \p_b \psi+\frac12M^2  \psi^2\bigg\}\,,
\label{eq:Lhxi}
\ea
where raising and lowering of the indices is now performed
with respect to the flat Euclidean space metric,
$\delta\ab$, since we are working perturbatively. 
Note that we are using the notation
$(\hat{h}^{a b})^2 \equiv \hat{h}^{ac}\, \hat{h}_{c}^{\  b }$.

We now calculate the one-loop matter contribution to the $n$-point graviton scattering amplitudes.
For simplicity of notation, we define the scattering amplitudes 
\ba
\nonumber
\mathcal{A}^{ (n\, \rm{pt})}\equiv \mathcal{A}^{ (n\, \rm{pt})}_{a_1 b_1 \cdots a_n b_n}h^{a_1 b_1}\cdots h^{a_n b_n}\ .
\ea
We start with the tadpole correction, using the dimensional regularization technique, which enables us to capture the running of the parameters of the theory. 

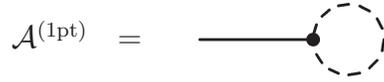
\begin{figure}[!htb]\vspace{20pt}
\begin{center}
$\mathcal{A}^{\rm (1pt)}\ \ =$\hspace{10pt}
\begin{fmffile}{Scattering2}
\parbox{50mm}{\begin{fmfgraph*}(70,30)
	            \fmfleft{i1}
	            \fmfright{o1}
                \fmflabel{$h\mn$}{i1}
                \fmf{plain}{i1,v1}
                \fmfdot{v1}
                \fmf{dashes,left=1,tension=0.8,label=$\psi$}{v1,o1}
                \fmf{dashes,left=1,tension=0.8}{o1,v1}
\end{fmfgraph*}}
\end{fmffile}
\end{center}
\caption{Contribution to the graviton tadpole from a matter loop.
Dashes denote the matter field propagator,
whereas solid lines denote the graviton.
This convention will be adopted throughout the paper.}
\label{Feynman_diagrams_1}
\end{figure}

\para{\bf Tadpole}
At one-loop,
the scalar field contributes to the graviton tadpole through the 3-vertex
$\hat h^{ab}\p_a \psi \p_b \psi$ 
represented in Fig. \ref{Feynman_diagrams_1}.
Explicit calculation of this vertex gives
\ba
\mathcal{A}^{\rm (1pt)}&=&\intk \frac{\hat h^{ab}k_a k_b}{k^2+M^2}=\frac14  M^4 [\hat{h}] J_{M,1}\,,
\label{eq:1ptmatter}
\ea
where
\ba\label{J1}
J_{M,1}=\frac{1}{M^{4}}\int \frac{\d^4 k}{(2\pi)^4} \frac{k^{2}}{k^2+M^2}\,,
\ea
as explained in appendix~\ref{sec:appendixDimReg}.

\para{\bf Beyond the tadpole}
We quote the result for the corrected $n$-point function here, and refer to appendix \ref{app:matter} for all the details.
Up to the $4$-point function and using Eqs.
 \eqref{eq:1ptmatter},
\eqref{eq:2ptmatterapp}, \eqref{eq:3pft} and \eqref{eq:4pft},
the counterterms which ought to added to the original action \eqref{eq:Lhxi} organize themselves into
\ba
\label{LCT_2}
\L_{CT}&=&-\(\mathcal{A}^{\rm (1pt)}+\frac{1}{2!} \mathcal{A}^{\rm (2pt)}
+\frac 1{3!} \mathcal{A}^{\rm (3pt)} +\frac 1{4!} \mathcal{A}^{\rm (4pt)}\)\nn \\
&=&-\frac{M^4}{4} \bigg( [\hat{h}] +\frac{1}{2!} \big( [\hat{h}]^2 -[\hat{h}^2] \big) +\frac{1}{3!}
\big( [\hat{h}]^3 +2[\hat{h}^3]  -3 [\hat{h}] [\hat{h}^2]\big)  \nn \\
& & \hspace*{1cm}+\frac{1}{4!}
\big( [\hat{h}]^4 -6[\hat{h}^4]  -6 [\hat{h}]^2 [\hat{h}^2] +3[\hat{h}^2]^2 +8 [\hat{h}] [\hat{h}^3]  \big)   \bigg)
J_{M,1}  \label{eq:expansiondet} \,. \\
&=&-\frac{M^4}{4} \sqrt{g} \ J_{M,1} \ \ . 
\label{eq:CC_contribution}
\ea
Note that the last line is only technically correct
if we include the zero-point function, which we can do (it is a vacuum bubble).
We conclude that the matter loops renormalize
the cosmological constant,
which is the only potential term one can obtain from integrating out matter loops, 
in agreement with the findings of Refs.~\cite{tHooft:1974bx,Park:2010rp}.
Importantly, matter loops do not affect the structure of the graviton potential.

\para{\bf Higher $n$-point functions}
From the one-loop effective action argument, we know that all the $n$-point functions will receive contributions
which will eventually repackage into the normalization of the cosmological constant. Seeing this explicitly at the perturbative level is nevertheless far less trivial, but  we give a heuristic argument here.
Taking the metric defined in Eq.~\eqref{eq:defmetric}, the expansion of the determinant of the metric to quartic order in $\hat{h}$ as given in \eqref{eq:expansiondet} is, in fact, \emph{exact}.
The finite nature of the running of the cosmological constant in \eqref{eq:expansiondet} is therefore no accident.

\begin{figure}[!htb]
\begin{center}
$\mathcal{A}^{\rm (5pt)}\ \ =$\hspace{15pt}
\begin{fmffile}{Scattering5pt}
\parbox{20mm}{\subfloat[]{\begin{fmfgraph*}(50,50)
	            \fmfsurround{i1,i2,i3,i4,i5}
                \fmf{plain}{i1,v1}
                \fmf{plain}{i2,v2}
                \fmf{plain}{i3,v3}
                \fmf{plain}{i4,v4}
                \fmf{plain}{i5,v5}
                \fmfdot{v1,v2,v3,v4,v5}
                \fmf{dashes,right=0.5,tension=0.9}{v1,v2,v3,v4,v5,v1}
\end{fmfgraph*}}}+
\parbox{20mm}{\subfloat[]{\begin{fmfgraph*}(50,50)
	            \fmfsurround{i1,i2,i3,i4,i5}
                \fmf{plain}{i1,v1}
                \fmf{plain}{i2,v2}
                \fmf{plain}{i3,v2}
                \fmf{plain}{i4,v3}
                \fmf{plain}{i5,v3}
                \fmfdot{v1,v2,v3}
                \fmf{dashes,right=0.6,tension=0.5}{v1,v2,v3,v1}
\end{fmfgraph*}}}+
\parbox{20mm}{\subfloat[]{\begin{fmfgraph*}(50,50)
	            \fmfsurround{i1,i2,i3,i4,i5}
                \fmf{plain}{i1,v1}
                \fmf{plain}{i2,v1}
                \fmf{plain}{i3,v1}
                \fmf{plain}{i4,v2}
                \fmf{plain}{i5,v2}
                \fmfdot{v1,v2}
                \fmf{dashes,left=0.7,tension=0.5}{v1,v2,v1}
\end{fmfgraph*}}}+
\parbox{20mm}{\subfloat[]{\begin{fmfgraph*}(50,50)
	            \fmfsurround{i1,i2,i3,i4,i5}
                \fmf{plain}{i1,v1}
                \fmf{plain}{i2,v2}
                \fmf{plain}{i3,v3}
                \fmf{plain}{i4,v3}
                \fmf{plain}{i5,v3}
                \fmfdot{v1,v2,v3}
                \fmf{dashes,right=0.5,tension=0.4}{v1,v2,v3,v1}
\end{fmfgraph*}}}+
\parbox{20mm}{\subfloat[]{\begin{fmfgraph*}(50,50)
	            \fmfsurround{i1,i2,i3,i4,i5}
                \fmf{plain}{i1,v1}
                \fmf{plain}{i2,v2}
                \fmf{plain}{i3,v3}
                \fmf{plain}{i4,v4}
                \fmf{plain}{i5,v4}
                \fmfdot{v1,v2,v3,v4}
                \fmf{dashes,right=0.4,tension=0.9}{v1,v2,v3,v4,v1}
\end{fmfgraph*}}}
\vspace*{0.5cm}\\
+
\parbox{20mm}{\subfloat[]{\begin{fmfgraph*}(50,50)
	            \fmfsurround{i1,i2,i3,i4,i5}
                \fmf{plain}{i1,v1}
                \fmf{plain}{i2,v2}
                \fmf{plain}{i3,v2}
                \fmf{plain}{i4,v2}
                \fmf{plain}{i5,v2}
                \fmfdot{v1,v2}
                \fmf{dashes,left=0.7,tension=0.4}{v1,v2,v1}
\end{fmfgraph*}}}+
\parbox{20mm}{\subfloat[]{\begin{fmfgraph*}(50,50)
	            \fmfsurround{i1,i2,i3,i4,i5}
                \fmf{plain}{i1,v1}
                \fmf{plain}{i2,v1}
                \fmf{plain}{i3,v1}
                \fmf{plain}{i4,v1}
                \fmf{plain}{i5,v1}
                \fmfdot{v1}
                \fmf{dashes,left=0.7,tension=0.8}{v1,v1}
\end{fmfgraph*}} }\\[20pt]
\end{fmffile}
\end{center}
\caption{One-loop contributions to the 5-point function.}
\label{Feynman_diagrams_5pt}
\end{figure}
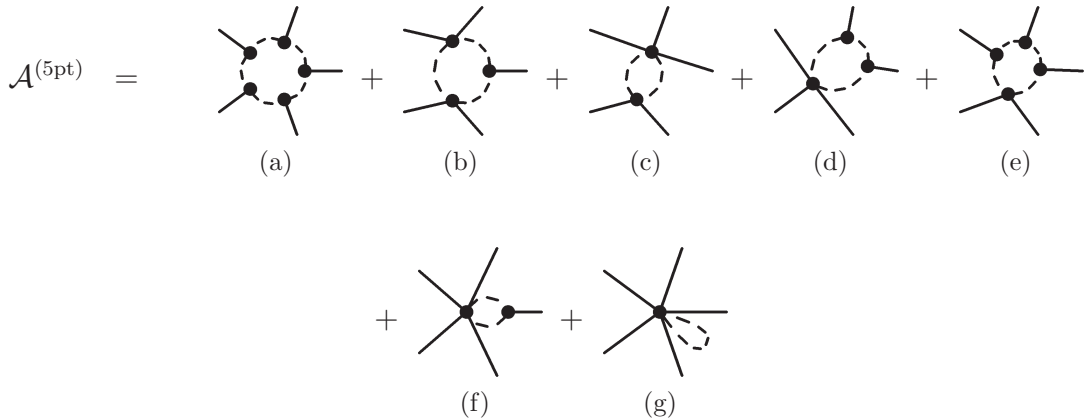

To show we would arrive at the same conclusion by explicit computation, consider the one-loop
correction to the 5-point function. In this case, there are five Feynman diagrams which contribute
at the same order for the quantum corrections, depicted in Fig. \ref{Feynman_diagrams_5pt}.
\noindent
From the interactions in the Euclidean action \eqref{eq:Lhxi}, we find
\ba
\mathcal{A}^{\rm (5pt)} &=& \frac{M^4}{4} \Bigg(
[\hat{h}]^5 +6 [\hat{h}^5] -\frac{15}{2} [\hat{h}] [\hat{h}^4] +5 [\hat{h}^3] [\hat{h}]^2
-5 [\hat{h}^3] [\hat{h}^2] +\frac{15}{4} [\hat{h}] [\hat{h}^2]^2  -\frac{5}{2} [\hat{h}^2] [\hat{h}]^3
\Bigg) \ J_{M,1}\nn\\
&\equiv&0 \,,\nn
\ea
which vanishes identically in four dimensions, as noted in Ref.~\cite{deRham:2010ik}. We can proceed in a similar manner to show that the same will be true for all the $n$-point functions, with $n>5$. This supports the consistency of the formalism introduced in \eqref{eq:defmetric} and explicitly agrees with the findings for GR as well as with the direct computation of the one-loop effective action.

Having shown the quantum stability of the massive gravity potential
at one-loop, one can see that the same remains true for any number of loops
provided there are no virtual gravitons running in the internal lines.

\section{Graviton loops}
\label{sec:GravitonLoops}

In the previous section we have studied in detail the quantum corrections
to the potential for massive gravity  arising from matter running in loops. We concluded that these quantum corrections could be resummed and interpreted as the renormalization of the cosmological constant. Therefore, we have shown that such loops are completely harmless to the special structure of the
ghost-free interaction potential.

Now we push this analysis forward by studying quantum corrections
from graviton loops. We start by considering
one-loop diagrams, and since we are interested in the IR limit of the theory, we set the external momenta to zero, as before.
We will again focus on the running of the interaction couplings, and thus apply dimensional regularization.

Based on studies within the decoupling limit \cite{deRham:2012ew},
we expect the quantum corrections to the graviton mass to scale as
$\delta m^2 \sim m^4/\mpl^2\sim 10^{-120} m^2$.
Even though such corrections are parametrically small, a potential problem arises if they
detune the structure of the interaction potential. If this happens,
ghosts arising at a scale much smaller than the Planck mass could in general plague the
theory, rendering it unstable against quantum corrections.
To show how such corrections could arise, we 
organise the loop diagrams in powers of the free parameters $\tilde{\alpha}_3$ and $\tilde{\alpha}_4$ of Eq.~\eqref{eq:Lvielbein}.

\subsection{Renormalization of the interactions}

We start by studying the quantum corrections arising at the linear order in the potential parameters $\tilde{\alpha}_{2,3,4}$. Since $\mathcal{\tilde{U}}_2$ is precisely quadratic in $h$, it leads to no corrections.
Next we focus on $\mathcal{\tilde{U}}_3$ in Eq.~\eqref{eq:Lvielbein}, which is cubic in $h$ and therefore can in principle renormalize the tadpole at one-loop. The tadpole contribution yields
\ba\label{Tadpole_alpha3}
\mathcal{A}^{\rm (1pt,3vt)} = -\frac58\tilde{\alpha}_3 \frac{m^4}{\mpl} [h] J_{m,1}\, ,
\ea
which on its own is harmless (this would correspond to the potential $\tilde{\mathcal{U}}_1$ which we have not included in \eqref{eq:Lvielbein},
 but which is also ghost-free \cite{Hassan:2011hr}). 
The last potential term $\mathcal{\tilde{U}}_4$ is quartic in $h$, as shown in \eqref{eq:U4h}.
This interaction vertex leads to quantum corrections to the 2-point function as shown in Fig.~\ref{1loopalpha4},
\ba
\mathcal{A}^{\rm (2pt,4vt)} = \mathcal{A}^{(2,4)}_{abcd} \,
h^{ab}h^{cd}
= 5 \tilde{\alpha}_4 \, \frac{m^4}{\mpl^2}  \([h^2]-[h]^2\) J_{m,1}\hspace{5pt}\propto \hspace{5pt} \tilde{\mathcal{U}}_2(h)\,,
\ea
where we have applied dimensional regularization with $J_{m,1}$
given in Eq.~\eqref{eq:Jn}.
This is nothing else but  the Fierz--Pauli structure, which is ghost-free by construction \cite{Fierz:1939ix}.\footnote{At the quadratic level,  the Fierz--Pauli term is undistinguishable from the ghost-free potential term $\tilde{\mathcal{U}}_2$.}

\begin{figure}[!htb]
\begin{center}
$\mathcal{A}^{\rm (2pt)}\ \ =$\hspace{15pt}
\begin{fmffile}{Scattering2grav}
\parbox{30mm}{\begin{fmfgraph*}(50,50)
                  \fmfsurround{i1,i2}
                \fmf{plain}{i1,v1}
                \fmf{plain}{i2,v1}
                \fmfdot{v1}
                \fmf{plain,right=0.5,tension=0.6}{v1,v1}
\end{fmfgraph*}}
\end{fmffile}
\end{center}
\caption{One-loop contribution to the
2-point correlation function from a graviton internal line.}
\label{1loopalpha4}
\end{figure}

\noindent Quadratic and other higher order corrections in $\tilde{\alpha}_3$ and $\tilde{\alpha}_4$ on the other hand are less trivial.

We will see however, that this optimistic
result will not prevail for other corrections,
which will induce the detuning of the interaction potential
structure in Eqs.~\eqref{eq:U2h}--\eqref{eq:U4h}.
To see this we turn to the interactions coming from the Einstein--Hilbert term.
Given \eqref{eq:defmetric} we can write the Einstein--Hilbert term as
\ba
\label{EH_term}
-\dfrac{\mpl^2}{2} \sqrt{g}R=h^{\alpha\beta} \Ein^{\mu\nu}_{\ \ \alpha\beta}
\, h\mn
+ \frac 1 \mpl h (\p h)^2+ \frac 1{\mpl^2} h^2 (\p h)^2 + \cdots \,,
\label{eq:EHterm}
\ea
where $ \Ein^{\mu\nu}_{\ \ \alpha\beta} $ is the usual Lichnerowicz operator written explicitly in Eq. \eqref{eq:Lichnerowicz}.

Contrary to the potential in \eqref{eq:U2h}--\eqref{eq:U4h},
Eq. \eqref{eq:EHterm}
contains an infinite numbers of interactions in $h$.
Taking, for example, the third and quartic order interactions from the Einstein--Hilbert term,
we find they do not generate any radiative correction to the tadpole,
$\mathcal{A}^{\rm(1pt,3vt)}_{\rm EH}=0$,
but they do contribute to the 2-point function as follows
\begin{equation}\label{EH3rd}
\mathcal{A}^{\rm (2pt)}_{\rm EH}=\frac{35}{12}\frac{m^4}{\mpl^2}(4[h^2]-[h]^2)J_{m,1}.
\end{equation}
The result above also does not preserve the Fierz--Pauli structure and is thus potentially dangerous.

\subsection{Detuning of the potential structure}
\label{sec:detuning}
From the above we conclude that quantum corrections from graviton loops can
spoil the structure of the ghost-free potential of massive gravity
required at the classical level to avoid propagating ghosts.
Interestingly, this detuning does not arise from the potential interactions at leading order in the parameters $\tilde{\alpha}_{3,4}$ but does arise from the kinetic Einstein--Hilbert term. 
Symbolically, the detuning of the potential occurs at the scale
\ba
\label{QC}
\L_{\rm qc}\sim \frac{m^4}{\mpl^n} \ h^n\,,
\ea
where $m$ is the graviton mass.
When working around a given background for $h=\bar h$ (which can include the helicity-0 mode, $\pi$),
this leads to a contribution at quadratic order which does \emph{not} satisfy the Fierz--Pauli structure,
\ba
\L_{\rm qc,\ \bar h}\sim \frac{m^4\bar{h}^{n-2}}{\mpl^n}\ h^2\, .
\ea
Reintroducing the canonically normalized helicity-0 mode as $h_{\mu\nu}=\p_\mu \p_\nu \pi/m^2$, this implies a ghost for the helicity-0 mode
\ba
\L_{\rm qc,\ \bar h}\sim \frac{\bar{h}^{n-2}}{\mpl^n}\ (\p^2 \pi)^2\sim \frac{1}{m_{\textrm{ghost}}^2}\ (\p^2 \pi)^2\, \hspace{20pt}{\rm with}\hspace{20pt}
m_{\textrm{ghost}}=\(\frac{\mpl}{\bar h}\)^{n/2}\bar h\, .
\ea
For interactions with $n\ge 3$, the mass of the ghost, $m_{\textrm{ghost}}$,
can be made arbitrarily \emph{small} by switching on an arbitrarily \emph{large} background configuration for $\bar h$.
This is clearly a problem since \emph{large} backgrounds ($\bar h \gtrsim \mpl$, or alternatively $\p^2 \bar \pi \gtrsim \Lambda_3^3$) are important for the Vainshtein mechanism \cite{Vainshtein:1972sx} to work and yet they can spoil the stability of the theory.

\section{One-loop effective action}
\label{sec:1PI}
We have shown that quantum corrections originated both from the potential as well as from the Einstein--Hilbert term in general destabilize the ghost-free interactions of massive gravity. This happens in a way which cannot be accounted for by either a renormalization of the coefficients of the ghost-free mass terms, or by a cosmological constant. This detuning leads to a ghost whose mass can be made arbitrarily small if there is a sufficiently large background source. From the decoupling limit analysis, we know that it is always possible to make the background source large enough without going outside of the regime of the effective field theory since we have
\be
\bar h \sim \frac{1}{m^2}\partial \partial \bar \pi = \mpl \frac{1}{\Lambda_3^3}\partial \partial \bar \pi\, .
\ee

As an effective field theory we are allowed
to make $\partial \partial \pi  \gg \Lambda_3^3$ provided $\partial^3 \pi /(1+\partial^2 \pi/\Lambda_3^3 ) \ll \Lambda_3^4$. In other words, as long as derivatives of the background $\bar h$ are small $\partial \ll \Lambda$, the magnitude of the background may be large $\bar h \gg \mpl$. 

The resolution of this problem in this case is that one also needs to take into account the redressing of the operators in the interaction potential. In this section we will investigate how the Vainshtein mechanism operates in protecting the effective action from the appearance of dangerous ghosts below the Planck scale.

\subsection{Form of the potential at one-loop}\label{subsec:olea}

Our previous approach involved explicit calculation of loop diagrams to evaluate the quantum corrections to the
massive gravity potential. Here, we shall focus on the formalism of the one-loop effective action
to confirm the destabilization result and provide a more generic argument. 
Since the quadratic potential $\tilde{\mathcal{U}}_2$ in \eqref{eq:U2h} has no non-linear interactions, we can take it as our sole potential term and consider all the graviton self-interactions arising from the Einstein--Hilbert term in Eq. \eqref{eq:EHterm}.
For simplicity, and without loss of generality, we therefore consider in what follows the specific theory of massive gravity
\ba
\L=-\dfrac{\mpl^2}{2} \sqrt{g}R - \frac 14 \mpl^2 m^2 \tilde{\mathcal{U}}_2[h]\,.
\label{eq:1lea}
\ea

We start by splitting the 
field $h\mn$
into a constant background $\bar h\mn$ and a
perturbation $\delta h\mn(x)$ which, in the language of the previous sections,
 will be the field running in the loops. We thus write
 $h\mn(x)=\bar h\mn+\delta h\mn(x)$.
 Up to quadratic order in the perturbation $\delta h$
\ba
\L &=& \frac 12 \delta h_{\alpha \beta}\(\Ein^{\mu\nu \alpha\beta} +m^2 \(\delta^{\mu\alpha} \delta^{\nu \beta}
- \delta^{\mu \nu} \delta^{\alpha \beta}\) \)\delta h_{\mu\nu} +\( \frac{1}{\mpl}\bar h  + \frac{1}{\mpl^2}\bar h^2 + \cdots\) (\p \delta h)^2\nn \\
&=& \frac 12 \delta h_{\alpha \beta} \(G^{-1 \ \mu\nu\alpha\beta}+M^{\mu\nu\alpha\beta}(\bar h)\)\delta h^{\mu\nu}\, \equiv
 \frac{1}{2} \delta h_{\alpha \beta}  \, \tilde{M}^{\mu\nu\alpha\beta}\,
 \delta h_{\mu\nu}\,,
 \label{eq:Leff1loop}
\ea
where $M^{\mu\nu\alpha\beta}(\bar h)= \(\frac 1 \mpl \bar h + \frac 1{\mpl^2} \bar h^2+\cdots \)\p^2 $
 symbolizes all the interactions in the Einstein--Hilbert term.
 $G^{-1}$ is the inverse of the massive graviton propagator.
 Following the same analysis of section \ref{subsec:oneloopea},
the one-loop effective action is then given by
\ba
\L_{\rm eff}&=&-\frac 12  \log \det  \(\frac{1}{\mu^2} \left\{G^{-1\ \mu\nu\alpha\beta}+M^{\mu\nu\alpha\beta}(\bar h)\right\}\)\nonumber \\
& \supseteq & -\frac{1}{2\mu^2} \(\int\frac{\d^4k}{(2\pi)^4}  \frac{f_{\mu\nu\alpha\beta}M^{\mu\nu\alpha\beta}}{\propm}
- \frac 12  \int \frac{\d^4k}{(2\pi)^4} \frac{f_{\mu\nu\alpha\beta}M^{\mu\nu a b } f_{abcd}M^{cd \alpha\beta}}{\propm^2} +\cdots   \)\, \ \,,
\label{eq:olearedressed}
\ea
where $M^{\mu\nu\alpha\beta}(\bar{h})$ is expanded in Fourier space and depends
explicitly on a derivative structure (and so,
on the momentum $k$). Here
$\mu$ denotes again a renormalization scale, which ought to be
introduced as a consequence of the renormalization procedure, and
to preserve the dimensional analysis.
Eq. \eqref{eq:olearedressed} sources an
effective potential which goes as $m^4 \mathcal{F}(\bar h / \mpl)J_{1,m} $ where $\mathcal{F}$
denotes an infinite series in powers of $\bar h$.
This result implies a running of the effective potential.

To gain some insight on the form of this effective potential,
we focus on the specific case of a conformally flat background where $\bar h\mn= \lambda \delta\mn$, for some real-valued $\lambda$.
It follows\footnote{This can be seen more explicitly, by writing the operator $M$ in terms of the background metric $\bar{\gamma}\ab$, recalling that $\bar g\ab=\bar{\gamma}_{ac}\bar{\gamma}_{bd}\delta^{cd}$ and the metric $g\ab$ is given in terms of $\bar{\gamma}\ab$ and the field fluctuation $h\ab$ as in \eqref{vielbeinForm}. Then it follows that symbolically,
\ba
M^{abcd}\sim (\delta^a_\mu\bar{\gamma}_{\nu \rho}\delta^{\rho b})(\delta ^c_\alpha \bar{\gamma}_{\beta \sigma}\delta^{\sigma d})
( \sqrt{\bar{g}} \, \bar{g}^{\mu\nu}\bar{g}^{\alpha \beta}
\bar{g}^{\bar{\gamma} \delta}\p_\gamma \p_\delta)\,,
\ea
where the two first terms in bracket arise from the transition to the `vielbein-inspired' metric fluctuation and the last term is what would have been otherwise the standard linearized Einstein--Hilbert term on a constant background metric $\bar{g}\ab$.
Written in this form, $M$ is manifestly conformally invariant.}
\ba
M^{\mu\nu\alpha\beta}(\bar h\mn=\lambda \delta\mn)=0\,,
\ea
which means that in this case all interactions are lost
and \eqref{eq:olearedressed} also vanishes.
This implies that the effective potential for a generic $\bar h\mn$ has to be of the form
\ba
\L_{\rm eff}= c_1 \([\bar h]^2-4[\bar h^2]\)+\(c_2[\bar h]^3+c_3[\bar h^2][\bar h]-(16 c_2+4c_3)[\bar h^3] \)+\cdots\,,
\label{eq:Leffd}
\ea
for some coefficients $c_1$, $c_2$ and $c_3$. The explicit form of these coefficients can be read off by computing specific Feynman diagrams corresponding to the Einstein--Hilbert interactions, or by considering a more general background metric $\bar h\mn$. For instance, $c_1$ corresponds to the coefficient in Eq.~\eqref{EH3rd}, $c_1=\frac{35}{12}\frac{m^4}{\mpl^2}$.
It is apparent that this structure is very different from that of the ghost-free potential of Eqs.
\eqref{eq:U2h}--\eqref{eq:U4h}. This confirms the results obtained in the previous sections.

\subsection{Mass of the ghost}
At what scale does this running arise?
Let us first concentrate on the quadratic term in \eqref{eq:Leffd}.
Since the helicity-0 mode $\pi$ enters as $h\mn= \p_\mu\p_\nu \pi/m^2$,
that term would lead to a correction of the form
\ba
\L_{\rm eff}^{(2)}=\frac{m^4}{\mpl^2}\([\bar h]^2-4[\bar h^2]\) J_{m,1} \sim \frac{1}{\mpl^2}\(\Box \pi\)^2 \ln(m^2/\mu^2)\, .
\ea
This would excite a ghost at the Planck scale.
Hence this contribution on its own is harmless.
Next 
we consider the effect of the cubic interactions,
\ba
\L_{\rm eff}^{(3)}=\frac{m^4}{\mpl^3} [\bar h]^3 J_1\sim \frac{1}{\mpl^3 m^2}
\(\Box \pi\)^3 \ln(m^2/\mu^2)\,.
\label{eq:beforeV1}
\ea
We now elaborate on the general argument mentioned in section \ref{sec:detuning}. 
We take a background configuration for $\pi$ wich is above the scale
$\Lambda_3=(\mpl m^2)^{1/3}$ for the Vainshtein mechanism to work. 
This will induce a splitting
of the helicity-0 mode $\p^2 \pi = \p^2 \bar \pi + \p^2 \delta \pi$,
 with $\p^2 \bar \pi \sim \mpl m^2/\kappa$ and $\kappa<1$. Then the operator in \eqref{eq:beforeV1} could lead to a ghost at a scale much lower than the Planck scale,
\ba
\L_{\rm eff}^{(3)}\sim \frac{1}{\mpl^2 \, \kappa}\(\Box \pi\)^2 \ln(m^2/\mu^2)\,.
\label{eq:beforeV2}
\ea
Thus by turning on a \emph{large} background, thereby making $\kappa$ \emph{smaller},
the scale at which the ghost arises becomes smaller and smaller, and
eventually comparable to $\Lambda_3$ itself.
This renders the theory unstable, as argued in section \ref{sec:detuning}.
 However, by assuming a large background we also need to understand its effect on the original
 operators via the Vainshtein mechanism \cite{Vainshtein:1972sx}.

\subsection{Vainshtein mechanism at the level of the one-loop 1PI}

The formalism of the one-loop effective action makes the Vainshtein mechanism particularly transparent
as far as the redressing of the interaction potential is concerned.
We further split the field $\bar h$ into a large background configuration $\bar{\bar{h}}\mn$ and a perturbation
$\tilde h\mn\sim \p_\mu\p_\nu \pi/m^2$,
such that $\bar{h}\mn=\bar{\bar{h}}\mn+\tilde h\mn$.
Since $\bar{\bar{h}}\mn$ satisfies the equations of motion we have
$\tilde{M}'(\bar{h})|_{\bar{\bar h}}=0$,
with $\tilde{M}$ defined as in Eq. \eqref{eq:Leff1loop}.
We proceed as before and expand the one-loop effective action up to second order in $\tilde h\mn$, as follows
\ba
\L_{\rm eff}&=&-\frac 12 \Tr \log \( \frac{1}{\mu^2}\tilde{M}^{\mu\nu\alpha\beta}(\bar h)\) \nonumber \\
&\supseteq&-\frac 12 \Tr \, \frac{G_{\mu\nu\alpha\beta}\tilde{M}''^{\mu\nu\alpha\beta}(\bar{\bar{h}})  }{G_{\mu\nu\alpha\beta} \tilde{M}^{\mu\nu\alpha\beta}(\bar{\bar{h}})} \, \tilde h ^2   \nonumber  \\
&\sim& \frac{m^4}{\mpl^2} \frac{\tilde{M}''(\bar{\bar{h}})}{\tilde{M}(\bar{\bar{h}})} [\tilde h]^2
\sim \frac{\Xi}{\mpl^2} \ (\Box \pi)^2 \,,
\ea
where the last line is symbolic,\footnote{In particular, by dimensional analysis,
one should think of the schematic form for the effective Lagrangian
 as containing a factor of $1/\mu^2$, where
$\mu$ carries units of $[\text{mass}]$.}
and for simplicity of notation we have denoted the combination
$\Xi \equiv \tilde{M}''(\bar{\bar{h}})/\tilde{M}(\bar{\bar{h}}) $.
It follows that the mass of the ghost is
$m^2_{\text{ghost}}=\Xi^{-1} \mpl^2 $.
Provided we can show that $\Xi \lesssim 1$,
the ghost will arise at least at the Planck scale and
the theory will always be under control.

\para{\bf Redressing the one-loop effective action}
The basis of the argument goes as follows. As explained previously, the Vainshtein mechanism relies on the fact that the background configuration can be large, and thus  $\tilde{M}''$ can in principle be large, which in turn can make $\Xi$ large and lower the mass of the ghost.
However, as we shall see in what follows, configurations with large $M''$ automatically lead to large $M$ as well.
This implies that $\Xi$ is always bounded $\Xi \lesssim 1$, and the mass of the ghost induced by the detuning/destabilization of the potential from quantum corrections is always at the Planck scale or beyond.\footnote{The only way to prevent $\Xi$ from being $\lesssim 1$ is to consider a region of space where some eigenvalues of the metric itself vanish, which would be for instance the case at the horizon of a black hole. However as explained in \cite{Deffayet:2011rh,Koyama:2011xz,Koyama:2011yg,Berezhiani:2011mt}, in massive gravity these are no longer coordinate singularities, but rather real singularities. In massive gravity, black hole
solutions ought to be expressed in such a way that the eigenvalues of the metric never reach zero apart at the singularity itself. Thus we do not need to worry about such configurations here (which would correspond to $\lambda=1$ in what follows).}

\para{\bf Computing the mass of the ghost}To make contact with an
explicit calculation we choose, for
simplicity, the following background metric $\bar g\ab=\text{diag}\{\lambda_0^2, \lambda_1^2, \lambda_1^2, \lambda_1^2\}=\bar{\gamma}\ab^2$,
and compute the possible combinations
appearing in $\Xi$. We define
\begin{equation}
\frac{\partial^2 \tilde{M}/\partial \bar{\bar{h}}_{\alpha\beta} \partial \bar{\bar{h}}_{\gamma\delta}}{\tilde{M}}\
\equiv \Xi_{(\alpha\beta,\gamma\delta)}\,,  \ \
\end{equation}
and use units for which $\mpl^2=1$.

In general, the components of $\tilde{M}$ can be split into three categories.
First, some components do not depend on the background, 
and are thus explicitly \emph{independent} of $\lambda_0$ and $\lambda_1$.
In this case $\tilde{M}''=\Xi=0$ trivially.
Second, other components are of the form
\begin{equation}
\tilde{M}\sim \frac{\lambda_0}{\lambda_1} \left( k_i k_j +m^2\right),
\label{eq:secondtildeM}
\end{equation}
where $i$ and $j$ are spatial indices. In this case
 $\Xi_{(00,00)}=0$,
 whereas $\Xi_{(ii,ii)} \sim \lambda_1^{-2} \lesssim 1$.

Finally the remaining category
contains terms which only depend on some power of the ratio of the components of the background metric, $\lambda_0/\lambda_1$ or $\lambda_1/\lambda_0$. The structure of the components of $\tilde{M}$ in this case is of the form\
\begin{equation}
\tilde{M}\sim \left(\frac{\lambda_1}{\lambda_0}\right)^p k_0^2
+\left(\frac{\lambda_0}{\lambda_1}\right)^q
\sum_i b_i k_i^2 \\ \  + \frac{\lambda_0}{\lambda_1} m^2,
\end{equation}
in which $i$ denotes a spatial component, $b_i$, $p$ and $q$ are integer numbers
such that $p, q \geq 1$.
The last term represents the (non-vanishing)
structure of the mass term.\footnote{The mass term
                                    can also arise in the form
                                    $(\lambda_1/\lambda_0) \, m^2$
                                    for the components $\tilde{M}^{00ii}$,
                                    but the conclusions hereafter
                                    remain unchanged.}
Let us work out the possible second derivatives in detail.
 On the one hand
\begin{equation}
\Xi_{(00,00)}
\sim
\dfrac{\left(\frac{\lambda_1}{\lambda_0}\right)^p  k_0^2 \, \lambda_0^{-2} +\left(\frac{\lambda_0}{\lambda_1}\right)^q \,  \lambda_0^{-2}\, \sum_i b_i k_i^2  }{\left(\frac{\lambda_1}{\lambda_0}\right)^p k_0^2
+\left(\frac{\lambda_0}{\lambda_1}\right)^p \frac{\lambda_0}{\lambda_1} \sum_i b_i k_i^2 +\textrm{mass term}}\ \,,
\end{equation}
where we have ignored factors of order unity to avoid clutter.
At first sight this result seems troublesome as it appears to be dependent on the choice of
background, and in particular on the hierarchy between $\lambda_0$ and
$\lambda_1$. We will however show that this is \emph{not} the case:
\begin{enumerate}[i.]
\item if $\lambda_0\sim\lambda_1\sim\lambda$, then  $\Xi_{(00,00)}\sim  \lambda^{-2}$. Since
we are interested in incorporating the Vainshtein effect, we shall consider the case
when $\lambda\gtrsim 1$, and thus
$\Xi\lesssim 1$.
\item if $\lambda_0 \gg \lambda_1 \gtrsim 1$, it follows
$\Xi \lesssim \lambda_0^{-2} \lesssim 1$;
the same holds true in the case $\lambda_1 \gg \lambda_0 \gtrsim 1$.
\end{enumerate}
On the other hand,
\begin{equation}
\Xi_{(jj,kk)}=
\Xi_{(jj,jj)}
\sim
\dfrac{\left(\frac{\lambda_1}{\lambda_0}\right)^p \, k_0^2 \,\lambda_1^{-2}+
 \left(\frac{\lambda_0}{\lambda_1}\right)^q \, \lambda_1^{-2}\, \sum_i b_i k_i^2
 + \frac{\lambda_0}{\lambda_1^3} m^2}{\left(\frac{\lambda_1}{\lambda_0}\right)^p  k_0^2
+\left(\frac{\lambda_0}{\lambda_1}\right)^q  \sum_i b_i k_i^2+
\frac{\lambda_0}{\lambda_1}m^2}\ \ .
\end{equation}
We repeat the previous analysis to show that,
regardless of the possible hierarchy between $\lambda_0$ and
$\lambda_1$, the quantum corrections will be parametrically small.
\begin{enumerate}[i.]
\item if $\lambda_0\sim\lambda_1\sim\lambda$, then
$\Xi_{(jj,jj)}\sim  \lambda^{-2}$, and hence $\Xi \lesssim 1$.
\item if either $\lambda_0 \gg \lambda_1 \gtrsim 1$
or $\lambda_1 \gg \lambda_0 \gtrsim 1$, then
$\Xi \lesssim \lambda_1^{-2} \ll 1$.
\end{enumerate}
Similar conclusions can be drawn for the `mixed' derivative $\Xi_{(00,ij)}$ or $\Xi_{(0i,0j)}$.

We have shown that, despite appearances, $\Xi\lesssim 1$ independently of the background and without loss of generality.
Whenever the Vainshtein mechanism is relevant, that is,
when $\lambda_0, \lambda_1 \gtrsim 1$,
the redressing of the operators
ensures that the mass of the ghost arises at least at the Planck scale.
We therefore conclude that at the one-loop level
the quantum corrections to the theory described by \eqref{eq:1lea} are under control.

\section{Conclusions}
\label{sec:conclusion}

In a theory of gravity both the mass and the structure of the graviton potential are fixed by phenomenological and theoretical constraints. While in GR this tuning is protected by covariance, such a symmetry is not present in massive gravity. Nevertheless, the `non-renormalization theorem' present in theories of  massive gravity implies that these tuning are technically natural \cite{deRham:2012ew,Nicolis:2004qq}, and hence do not rely on the same fine-tuning as for instance setting the cosmological constant to zero. In this manuscript we have explored the stability of the graviton potential further by looking at loops of matter and graviton, assuming a covariant coupling to matter (see for instance Ref.~\cite{deRham:2011qq} for a discussion of the natural coupling to matter and its stability).

When integrating out externally coupled matter fields, we have shown  explicitly that the only potential contribution to the one-loop effective action is a cosmological constant, and the special structure of the potential is thus unaffected by the matter fields at one-loop.

For graviton loops, on the other hand, the situation is more involved---they \emph{do} change the structure of the potential, but in a way which only becomes relevant at the Planck scale. 
Nevertheless, the Vainhstein mechanism that resolves the vDVZ discontinuity relies on a classical background configuration to exceed the Planck scale (\ie $g\mn -\delta\mn\gtrsim 1$), without going beyond the regime of validity of the theory. A na\"{i}ve perturbative estimate would suggest that on top of such large background configurations, the mass of the ghost could be 
lowered well below the Planck scale. However, this perturbative argument does not take into account the same Vainshtein mechanism that suppresses the vDVZ discontinuity in the first place. To be consistent we have therefore considered a non-perturbative background and have shown that the one-loop effective action is itself protected by a similar Vainshtein mechanism which prevents the mass of the ghost from falling below the Planck scale, even if the background configuration is large.

The simplicity of the results presented in this study rely on the fact that the coupling to matter is taken to be covariant and that at the one-loop level virtual gravitons and matter fields cannot mix. Thus at one-loop virtual matter fields remain unaware of the graviton mass. This feature is lost at higher loops where virtual graviton and matter fields start mixing.

Higher order loops are beyond the scope of this paper, but will be investigated in depth in Ref.~\cite{2loops}. In this follow-up study, we will show how a na\"{i}ve estimate would suggest that the two-loop graviton-matter mixing can lead to a detuning of the potential already at the scale $\mpl (m/M)^2$, where $m$ is the graviton mass and $M$ is the matter field mass.
If this were the case, a matter field with $M\sim\Lambda_3$ would already bring the mass of the ghost down to $\Lambda_3$ which would mean that the theory could never be taken seriously beyond this energy scale (or its redressed counterpart, when working on a non-trivial background). However, this estimate does not take into account the very special structure of the ghost-free 
graviton potential which is already manifest in its decoupling limit. Indeed, in ghost-free massive gravity the special form of the potential leads to interesting features when mixing matter and gravitons in the loops.

\begin{figure}[!htb]
\begin{center}
$\mathcal{A}_{\chi\chi}=$ \hspace{0.5cm}
\begin{fmffile}{ScalarField2pt_3}
\parbox{40mm}{\subfloat[]{\begin{fmfgraph*}(70,30)
	            \fmfleft{i1}
	            \fmfright{o1}
                \fmf{dashes}{i1,v1}
                \fmf{dashes}{v2,o1}
                \fmfdot{v1}
                \fmfdot{v2}
                \fmf{plain,left=1,tension=0.8}{v1,v2}
                \fmf{dashes,left=1,tension=0.}{v2,v1}
\end{fmfgraph*}}}
\hspace{-20pt}$+$\hspace{30pt}
\parbox{40mm}{\subfloat[]{\begin{fmfgraph*}(70,30)
	            \fmfsurround{i1,i2}
                \fmf{dashes}{i1,v1}
                \fmf{dashes}{i2,v1}
                \fmfdot{v1}
                \fmf{plain,right=0.5,tension=1.}{v1,v1}
\end{fmfgraph*}}}
\end{fmffile}
\end{center}
\caption{Contribution to the scalar field two-point function from graviton/matter loops.}
\label{Fig:ScalarField2pt}
\end{figure}
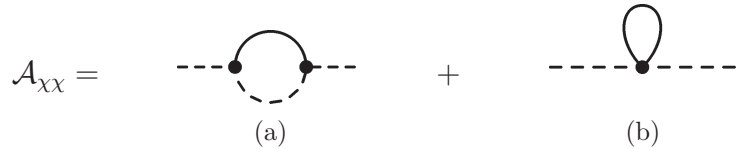

To give an idea of how this mixing between gravitons and matter arises, let us consider the one-loop contribution to the scalar field two-point function depicted in Fig.~\ref{Fig:ScalarField2pt} if the scalar field does not have any self-interactions. We take the external leg of the scalar field to be on-shell, \ie with momentum $q_a$ satisfying $\delta^{ab}q_a q_b+M^2=0$.\footnote{Technically, in Euclidean space this means that the momentum is complex, or one could go back to the Lorentzian space-time for the purposes of this calculation, but these issues are irrelevant for the current discussion. Moreover, note that
the on-shell condition is only being imposed for the external legs, and \emph{not} for internal lines.}

For the purpose of this discussion, it is more convenient to work in terms of the field $\chi$ directly rather than the redefined field $\psi$.
First, diagram (a) gives rise to a contribution proportional to
\ba
\mathcal{A}^{(a)}_{\chi\chi}\propto \frac{1}{\mpl^2}\int \d^4 k \frac{f^{(m)}_{abcd}(k)\, q^ap^bq^cp^d}{(k^2+m^2)(p^2+M^2)}\,,
\ea
where $k$ is the momentum of the virtual graviton running in the loop of diagram (a).
By momentum conservation, the momentum $p$ of the virtual field $\chi$ in the loop is then $p_a=q_a-k_a$. Applying the on-shell condition for the external legs, $q^2+M^2=0$, we find
\ba
\mathcal{A}^{(a)}_{\chi\chi}\propto \frac{1}{m^4\mpl^2}\int \d^4 k \, \((k.q)^2+m^2 q^2\)\equiv 0 \hspace{10pt}\text{in dimensional regularization}\,.
\ea
So this potentially `problematic' diagram that mixes virtual matter and gravitons (which could a priori scale as $m^{-4}$) leads to no running when the 
\emph{external} scalar field is on-shell. In other words, at most this diagram can only lead to a running of the wave-function
normalization and is thus harmless (in particular it does not affect the scalar field mass, nor does it change the `covariant' structure of the scalar field Lagrangian).

Second, we can also consider the contribution from a pure graviton-loop in diagram (b). Since only the graviton runs in that loop, the running of the scalar field mass arising from that diagram is at most $\delta M^2= \frac{m^2}{\mpl^2}M^2 \ll M^2$ and is therefore also harmless.
As a consequence we already see in this one-loop example how the mixing between the virtual graviton and scalar field in the loops keeps the structure of the matter action perfectly under control.

\acknowledgments

We would like to thank Michael Duff, Gregory Gabadadze, Kurt Hinterbichler, Rachel Rosen and Andrew J.~Tolley for useful discussions, and Matteo Fasiello for comments on an early draft of this paper. CdR is supported by Department of Energy grant DE-SC0009946. LH is supported by the Swiss National Science Foundation.
We acknowledge the use of the xAct package for
Mathematica \cite{Brizuela:2008ra, xAct}.

\appendix

\section{Dimensional regularization}
\label{sec:appendixDimReg}

For the one-loop diagrams we required the dimensional
regularization technique to obtain the quantum corrections.
A recurrent integral which appears in our calculations is of the form
\ba
J_{\tilde{m},n}=\frac{1}{\tilde{m}^{4}}\int \frac{\d^4 k}{(2\pi)^4} \frac{k^{2n}}{\(k^2+\tilde{m}^2\)^n}\,,
\label{eq:Jn}
\ea
where $\tilde{m}$ is a placeholder for whichever mass
appears in the propagator.
By symmetry we have
\ba
\frac{1}{\tilde{m}^{4}}\int \frac{\d^4 k}{(2\pi)^4} \frac{k^{2(n-j)}k_{\alpha_1} \cdots  k_{\alpha_{2j}}}{\(k^2+\tilde{m}^2\)^n}=\frac{1}{2^{j} (j+1)!} \delta_{\alpha_1 \cdots \alpha_{2j}}  J_{\tilde{m},n}\,,
\ea
with the generalized Kronecker symbol, 
\begin{equation}
\delta_{\alpha_1 \cdots \alpha_{2j}}=
\delta_{\alpha_1 \alpha_2} \delta_{\alpha_3 \cdots \alpha_{2j}}+
\Big( \{\alpha_2\} \leftrightarrow \{\alpha_3, \cdots , \alpha_{2j}\}\Big)\ .
\end{equation}
We also note that
\ba
J_{\tilde{m},n}=\frac{n(n+1)}{2} J_{\tilde{m},1}\,.
\ea
We do not need to express $J_{\tilde{m},1}$ explicitly in dimensional regularization, but can simply rely on these different relations to show how different diagrams repackage into a convenient form.
It suffices to know that $J_{\tilde{m},1}$ contains the logarithmic divergence in $\tilde{m}$, which represents the running in renormalization techniques.

\section{A closer look at the matter loops}
\label{app:matter}

In section \ref{subsec:renormLambda} we have calculated the quantum corrections
arising from matter loops to the graviton tadpole. For completeness, we collect
in this appendix the individual corrections
from each Feynman diagram corresponding to a given $n$-point function.

\para{\bf 2-point function}There are two Feynman diagrams which contribute
to the corrected 2-point function, which arise respectively from
the cubic and quartic interactions in the action \eqref{eq:Lhxi}.

\begin{figure}[!htb]
\begin{center}
$\mathcal{A}^{\rm (2pt)}\ \ =$\hspace{15pt}
\begin{fmffile}{Scattering2pt}
\parbox{20mm}{\subfloat[]{\begin{fmfgraph*}(50,50)
	            \fmfsurround{i1,i2}
                \fmf{plain}{i1,v1}
                \fmf{plain}{i2,v2}
                \fmfdot{v1,v2}
                \fmf{dashes,right=0.5,tension=0.2}{v1,v2,v1}
\end{fmfgraph*}}}+
\parbox{20mm}{\subfloat[]{\begin{fmfgraph*}(50,50)
	            \fmfsurround{i1,i2}
                \fmf{plain}{i1,v1}
                \fmf{plain}{i2,v1}
                \fmfdot{v1}
                \fmf{dashes,right=0.5,tension=0.7}{v1,v1}
\end{fmfgraph*}}} \\[10pt]
\end{fmffile}
\end{center}
\caption{Contribution to the graviton 2-point function from matter loops.}
\label{Feynman_diagrams_2pt}
\end{figure}

\noindent Evaluation of these one-loop diagrams shown in Fig. \ref{Feynman_diagrams_2pt}
gives
\ba
\mathcal{A}^{\rm (2pt)}_{\rm (a)}&=& 2 \hat{h}^{ab}\hat{h}^{cd} \intk  \frac{k_a k_b k_c k_d}{(k^2+M^2 )^2}
=\frac1{4} M^4\(2[\hat{h}^2]+[\hat{h}]^2\) J_{M,1}\,, \nn\\
\mathcal{A}^{\rm (2pt)}_{\rm (b)} &=&-3  \intk  \frac{(\hat{h}^2)^{ab}k_a k_{b} }{(k^2+M^2 )}
=- \frac{3}{4} \frac{M^4}{\mpl^2}\, [\hat{h}^2]\ J_{M,1} \,,
\nn
\ea
so that the total contribution to the 2-point function is 
\ba
\mathcal{A}^{\rm (2pt)}&=&\mathcal{A}^{\rm (2pt)}_{\rm (a)}+\mathcal{A}^{\rm (2pt)}_{\rm (b)}
=\frac1{4} M^4\big([\hat{h}]^2-[\hat{h}^2]\big) J_{M,1}\,.
\label{eq:2ptmatterapp}
\ea

\para{\bf  3-point function}
The three-point scattering amplitude will receive corrections from the diagrams
depicted in Fig. \ref{Feynman_diagrams_3pt}
\begin{figure}[!htb]\vspace{10pt}
\begin{center}
$\mathcal{A}^{\rm (3pt)}\ \ =$\hspace{15pt}
\begin{fmffile}{Scattering3pt}
\parbox{20mm}{\subfloat[]{\begin{fmfgraph*}(50,50)
	            \fmfsurround{i1,i2,i3}
                \fmf{plain}{i1,v1}
                \fmf{plain}{i2,v2}
                \fmf{plain}{i3,v3}
                \fmfdot{v1,v2,v3}
                \fmf{dashes,right=0.5,tension=0.4}{v1,v2,v3,v1}
\end{fmfgraph*}}}+
\parbox{20mm}{\subfloat[]{\begin{fmfgraph*}(50,50)
	            \fmfsurround{i1,i2,i3}
                \fmf{plain}{i1,v1}
                \fmf{plain}{i2,v2}
                \fmf{plain}{i3,v2}
                \fmfdot{v1,v2}
                \fmf{dashes,left=0.7,tension=0.4}{v1,v2,v1}
\end{fmfgraph*}}}+
\parbox{20mm}{\subfloat[]{\begin{fmfgraph*}(50,50)
	            \fmfsurround{i1,i2,i3}
                \fmf{plain}{i1,v1}
                \fmf{plain}{i2,v1}
                \fmf{plain}{i3,v1}
                \fmfdot{v1}
                \fmf{dashes,left=0.7,tension=0.7}{v1,v1}
\end{fmfgraph*}}} \\[20pt]
\end{fmffile}
\end{center}
\caption{One-loop contributions to the 3-point function.}
\label{Feynman_diagrams_3pt}
\end{figure}
which give the following contributions
\ba
\mathcal{A}^{\rm (3pt)}_{\rm(a)} &=& \frac{M^4}{4} \([\hat{h}]^3+6[\hat{h}][\hat{h}^2]+8[\hat{h}^3]\) J_{M,1}\,, \nn\\
\mathcal{A}^{\rm (3pt)}_{\rm(b)} &=& -\frac{9 M^4}{4} \([\hat{h}^2][\hat{h}]+2[\hat{h}^3]\) J_{M,1} \nn \,,\\
\mathcal{A}^{\rm (3pt)}_{\rm(c)} &=&  3 M^4  [\hat{h}^3]    \, J_{M,1} \ \nn .
\ea
We conclude the total 3-point function goes as
\ba
\mathcal{A}^{\rm (3pt)} &=& \frac{M^4}{4} \( 2[\hat{h}^3] +[\hat{h}]^3 -3 [\hat{h}] [\hat{h}^2] \)  \, J_{M,1} \, .
\label{eq:3pft}
\ea

\para{\bf  4-point function}
The Feynman diagrams contributing to the corrected 4-point function are
those in Fig. \ref{Feynman_diagrams_4pt}
\begin{figure}[!htb]\vspace{20pt}
\begin{center}
$\mathcal{A}^{\rm (4pt)}\ \ =$\hspace{15pt}
\begin{fmffile}{Scattering4pt}
\parbox{20mm}{\subfloat[]{\begin{fmfgraph*}(50,50)
	            \fmfsurround{i1,i2,i3,i4}
                \fmf{plain}{i1,v1}
                \fmf{plain}{i2,v2}
                \fmf{plain}{i3,v3}
                                \fmf{plain}{i4,v4}
                \fmfdot{v1,v2,v3,v4}
                \fmf{dashes,right=0.5,tension=0.4}{v1,v2,v3,v4,v1}
\end{fmfgraph*}}}+
\parbox{20mm}{\subfloat[]{\begin{fmfgraph*}(50,50)
	            \fmfsurround{i1,i2,i3,i4}
                \fmf{plain}{i1,v1}
                \fmf{plain}{i2,v1}
                \fmf{plain}{i3,v2}
                                \fmf{plain}{i4,v2}
                \fmfdot{v1,v2}
                \fmf{dashes,left=0.7,tension=0.4}{v1,v2,v1}
\end{fmfgraph*}}}+
\parbox{20mm}{\subfloat[]{\begin{fmfgraph*}(50,50)
	            \fmfsurround{i1,i2,i3,i4}
                \fmf{plain}{i1,v1}
                \fmf{plain}{i2,v1}
                \fmf{plain}{i3,v1}
                                \fmf{plain}{i4,v2}
                \fmfdot{v1,v2}
                \fmf{dashes,left=0.7,tension=0.3}{v1,v2,v1}
\end{fmfgraph*}}}+
\parbox{20mm}{\subfloat[]{\begin{fmfgraph*}(50,50)
	            \fmfsurround{i1,i2,i3,i4}
                \fmf{plain}{i1,v1}
                \fmf{plain}{i2,v1}
                \fmf{plain}{i3,v2}
                                \fmf{plain}{i4,v3}
                \fmfdot{v1,v2,v3}
                \fmf{dashes,right=0.7,tension=0.4}{v1,v2,v3,v1}
\end{fmfgraph*}}}+
                \parbox{20mm}
                {\subfloat[]{\begin{fmfgraph*}(50,50)
	            \fmfsurround{i1,i2,i3,i4}
                \fmf{plain}{i1,v1}
                \fmf{plain}{i2,v1}
                \fmf{plain}{i3,v1}
                                \fmf{plain}{i4,v1}
                \fmfdot{v1}
                \fmf{dashes,right=0.7,tension=0.7}{v1,v1}
\end{fmfgraph*}} }\\[20pt]
\end{fmffile}
\end{center}
\caption{One-loop contributions to the 4-point function.}
\label{Feynman_diagrams_4pt}
\end{figure}
and they give the following contributions
\ba
\mathcal{A}^{\rm (4pt)}_{\rm(a)} &=& M^4 \Big(12[\hat{h}^4]+8[\hat{h}][\hat{h}^3]+3[\hat{h}^2]^2
+3 [\hat{h}^2] [\hat{h}]^2 +\frac{1}{4} [\hat{h}]^4\Big) J_{M,1}\,, \nn \\
\mathcal{A}^{\rm (4pt)}_{\rm(b)} &=& \frac{27 M^4}{4} \([\hat{h}^2]^2+2[\hat{h}^4]\) J_{M,1} \,, \nn \\
\mathcal{A}^{\rm (4pt)}_{\rm(c)} &=&  12 M^4  \([\hat{h}^3] [\hat{h}] +2 [\hat{h}^4] \)   \, J_{M,1} \,, \nn\\
\mathcal{A}^{\rm (4pt)}_{\rm(d)} &=&  -\frac{9 M^4}{2} \( [\hat{h}^2] [\hat{h}]^2
+2 [\hat{h}^2]^2 +2 [\hat{h}] [\hat{h}^3] +8 [\hat{h}^4] +2 [\hat{h}][\hat{h}^3]\)  \, J_{M,1} \,,\nn \\
 \mathcal{A}^{\rm (4pt)}_{\rm(e)} &=&  -15 M^4 [\hat{h}^4] \, J_{M,1} \,, \nn
\ea
so that the total 4-point function is given by
\ba
\mathcal{A}^{\rm (4pt)} &=& \frac{M^4}{4} \( [\hat{h}]^4 -6  [\hat{h}^4] -6  [\hat{h}^2] [\hat{h}]^2
+3  [\hat{h}^2]^2  +8  [\hat{h}]   [\hat{h}^3]
\)\, J_{M,1} \, .
\label{eq:4pft}
\ea
 Eqs. \eqref{eq:2ptmatterapp}, \eqref{eq:3pft} and \eqref{eq:4pft} have the precise coefficients
 to produce a running of the cosmological constant, as shown in Eq. \eqref{eq:expansiondet}.

	\bibliographystyle{JHEPmodplain}
	\bibliography{references}

\end{document}